\documentclass[11pt, draft]{article}
\usepackage{calc}
\usepackage[charter]{mathdesign}
\usepackage{amsmath}
\usepackage{tikz}
\usepackage{bbding} 
\usepackage{caption}
\usepackage{booktabs} 
\usepackage{natbib} 
\usepackage{newunicodechar}
	\newunicodechar{“}{``}
	\newunicodechar{”}{''}
	\newunicodechar{’}{'}
	\newunicodechar{≤}{\le}
	\newunicodechar{≥}{\ge}
	\newunicodechar{µ}{\mu}
	\newunicodechar{𝛼}{\alpha}
	\newunicodechar{𝛽}{\beta}
	\newunicodechar{𝛾}{\gamma}
	\newunicodechar{𝜋}{\pi}
	\newunicodechar{𝜏}{\tau}
	\newunicodechar{𝜔}{\omega}
	\newunicodechar{𝜎}{\sigma}
	\newunicodechar{𝜅}{\kappa}
	\newunicodechar{𝜃}{\theta}
	\newunicodechar{𝜌}{\rho}
	\newunicodechar{𝛿}{\delta}
	\newunicodechar{𝜀}{\varepsilon}
	\newunicodechar{𝜁}{\zeta}
	\newunicodechar{∂}{\partial}
	\newunicodechar{𝛥}{\Delta}
	\newunicodechar{×}{\times}
	\newunicodechar{∑}{\sum}
	\newunicodechar{╟}{\left(}
	\newunicodechar{╢}{\right)}
	\newunicodechar{…}{\dots}
	\newunicodechar{⋯}{\cdots}
	\newunicodechar{ⅆ}{\mathop{}\mathrm{d}}
	\newunicodechar{ⅇ}{\mathrm{e}}
	\newunicodechar{╻}{_{t_n+𝛥t}}
	\newunicodechar{∙}{\cdot}
	\newunicodechar{–}{--}

\usepackage{siunitx}
\usepackage[paper=a4paper, margin=2.25cm, includefoot]{geometry}
\usepackage[english]{babel}

\title{Explicit solution to dynamic portfolio choice problem: the~continuous-time detour}

\date{February 16, 2015}

\author{François Legendre\thanks{Érudite, Université Paris-Est, and TEPP, \texttt{F.Legendre@u-pec.fr}.} \and Djibril Togola\thanks{\protect\raisebox{-.25ex}{\Envelope} Érudite, Université Paris-Est, \texttt{Djibril.Togola@u-pec.fr}.}}

\begin{document}
\lefthyphenmin 3 \righthyphenmin 3 

\maketitle

\begin{center}
\itshape
\end{center}

\bigskip

\begin{abstract}
\noindent 
This paper solves the dynamic portfolio choice problem. Using an explicit solution with a power utility, we construct a bridge between a continuous and discrete VAR model to assess portfolio sensitivities. We find, from a well analyzed example that the optimal allocation to stocks is particularly sensitive to Sharpe ratio.  Our quantitative analysis highlights that this sensitivity increases when the risk aversion decreases and/or when the time horizon increases. This finding explains the low accuracy of discrete numerical methods especially along the tails of the unconditional distribution of the state variable.
\end{abstract}

\bigskip

\textbf{Keywords}: Dynamic portfolio choice; Long-term investing; Time aggregation; Explicit solution; 

Numerical solution. 

\bigskip

\textbf{JEL Classification}: G11; G12.

\bigskip

\section*{Introduction}
Since at least \citet{Merton-1971}, many results on portfolio optimization problems have been obtained in a continuous time framework. It is still difficult to solve optimal portfolio problems when there is some degree of predictability in asset returns, \emph{i.e.} when the investment opportunities are time-varying. A great number of papers have proposed to use a VAR model to forecast returns and study its implication for the long-term portfolio choice problem. As a result the academic literature has followed two main directions. The first one relies on mathematical tools and establishes some explicit solutions (see among others \citet{Kim-Omberg-1996}, \citet{Liu-2007} and references therein). The second line of research consists to implement some challenging numerical methods. In fact, \cite{Barberis-2000} developed a discretization state space method which serves as a benchmark. \cite{Brandt-al.-2005}, \cite{Binsbergen-Brandt-2007}, \citet{Garlappi-Skoulakis-2009} among others use some sophisticated backward induction techniques and evaluate the accuracy of their results by comparing them to the discretization state space benchmark. However, all discrete numerical procedures approximate directly or indirectly a highly non linear value function and cannot explicitly separate the so-called \emph{hedging demand} from the so-called \emph{myopic demand}. \citet{Garlappi-Skoulakis-2011} provide a general discussion about approximations accuracy in discrete time.

This paper works in continuous time and uses the explicit solution for the portfolio choice problem, then constructs a bridge between continuous and discrete VAR model as in \citet{Campbell-al.-2004}. In fact, these authors provided evidence that there should exist minor discrepancies between results under discrete and continuous time models. Thus, numerical results that we derive from continuous time are indirectly comparable to those of \citet{Garlappi-Skoulakis-2009}. We show that, for large degrees of risk aversion and/or small horizon, when the state variable closes to its unconditional mean, the two numerical results are quite similar. Otherwise, results under our explicit solution in continuous time exhibit some discrepancies with \citet{Garlappi-Skoulakis-2009} when the risk aversion decreases and/or when the time horizon increases. We argue that this is due to large sensitivity of total demand to state variable (Sharpe ratio).

The paper is organized as follows. Section 1 exposes the way we map the continuous-time investment opportunity set and the discrete-time one. Section 2 gives some insights on the explicit solution for the long-term investor with CRRA preferences. Section 3 gives some numerical results based on~\citet{Brandt-al.-2005} example.
 
\section{Investment opportunity sets}

We first expose the model in a continuous-time framework and in a discrete-time framework to study the impact of a predictable component in stock returns. Next, we show how to recover continuous-time parameters that are consistent with discrete-time VAR estimates.

\subsection{Opportunity set in continuous time}

We start by assuming that two assets are available for the investor (\citet{Campbell-al.-2004} and \citet{Kim-Omberg-1996} among others). On the one hand, the riskless asset pays back a constant interest rate~$r$
\begin{equation}
\frac{ⅆ P^f_t}{P^f_t} = r ⅆt \label{eq:Pf}\\
\end{equation}
where~$P^f_t$ denotes the price of this asset at time~$t$. On the other hand, there is a risky asset whose price~$P_t$ satisfies the following diffusion process
\begin{equation}
\frac{ⅆ P_t}{P_t} = µ_t ⅆt + 𝜎 ⅆB_t^p \\
\end{equation}
where~$B_t^p$ denotes a scalar Brownian motion with zero drift and unit variance rate. The drift rate~$µ_t$ follows a diffusion process as well. It is supposed to be time-varying and state variable dependent. The volatility of the risky asset is assumed to be constant. This is not a strong assumption for the long-term investor (see \citet{Campbell-Viceira-2002}). Let~$X_t$ denote the Sharpe ratio i.e. the market price of risk/reward for buying/selling one unit of risky asset
\begin{equation}
X_t = \frac{µ_t - r}{𝜎}
\end{equation}
Then the Sharpe ratio is assumed to follow the usual ``Ornstein-Uhenbeck'' diffusion process
\begin{equation}
ⅆ X_t = 𝜅(𝜃 - X_t) ⅆt + 𝜁 ⅆB_t^x \quad 𝜅, 𝜃, 𝜁 > 0
\end{equation}
where~$B_t^x$ denotes another scalar Brownian motion with zero drift and unit variance rate. 
Parameters~$𝜃$ and~$𝜅$ denote respectively the unconditional mean and the mean reverting parameter of the Sharpe ratio~$X_t$. In fact, parameter~$𝜅$ reflects the rate by which the shocks on Sharpe ratio dissipate and then reverts towards its long-term mean~$𝜃$. Finally, parameter~$𝜁$ denotes the instantaneous volatility of Sharpe ratio. It controls the diffusion rate of the process. 

The above equations imply that instantaneous return on stocks~$ⅆ P_t/P_t$ follows a diffusion process whose drift is mean-reverting and whose innovations are correlated with those of the market price of risk itself, with the correlation coefficient~$𝜌$. Thus the following equations hold.
\begin{align}
ⅆ P_t/P_t & = (𝜎 X_t + r) ⅆ t + 𝜎 ⅆB_t^p \label{eq:cont1}\\
ⅆ X_t & = 𝜅(𝜃 - X_t) ⅆt + 𝜁 ⅆB_t^x \label{eq:cont2}
\end{align}
with~$ⅆB_t^p ⅆB_t^x = 𝜌 ⅆt$.
Equations~\eqref{eq:cont1} and~\eqref{eq:cont2} define a joint stochastic process in continuous time.

\subsection{Opportunity set in discrete time}

The standard model in discrete time is a restricted VAR(1) process which captures predictability of stocks returns (see \citet{Barberis-2000} for instance). We focus on the example analyzed in \citet{Brandt-al.-2005} that was reused in \citet{Binsbergen-Brandt-2007} and in \citet{Garlappi-Skoulakis-2009}. The log excess returns of the risky asset~$𝛥 \ln P_{t+1} - r^f$ are assumed to be predictable by the log dividend-to-price ratio~$z_t$ ($r^f$ denotes the risk-free rate and is equal to 6\% in annualized basis). The joint dynamics of these two variables are specified such that
\begin{align}
𝛥 \ln P_{t+1} - r^f & = a_r + b_r z_t + 𝜀^r_{t+1} \label{eq:disc1}\\
z_{t+1} & = a_z + b_z z_t + 𝜀^z_{t+1} \label{eq:disc2}
\end{align}
with
\begin{equation}
\begin{pmatrix}
𝜀^r_{t+1}\\𝜀^z_{t+1}
\end{pmatrix}
\sim N
\begin{bmatrix}
\begin{pmatrix}0\\0\end{pmatrix},
\begin{pmatrix}𝜎^2_r & 𝜎_{rz} \\ 𝜎_{rz} & 𝜎^2_z\end{pmatrix}
\end{bmatrix}
\end{equation}
In fact, \citet{Campbell-Shiller-1988} forcefully claim that, if returns are predictable, at least, the log dividend-to-price ratio should capture some of that predictability. A substantial long-standing empirical literature has documented many properties of these two regressions. \citet{Brandt-al.-2005} report the following estimated values (using the CRSP U.S. quarterly index from January 1986 to December 1995)
\[
b_r = \num{.060} \quad b_z = \num{.958} \quad \frac{𝜎_{rz}}{𝜎_r \, 𝜎_z} = \num{-.941}
\]
The returns are weakly predictable, the dividend yield is highly persistent and the shocks are strongly negatively related.

\subsection{Recovering continuous-time parameters from discrete-time VAR}

We closely follow \citet{Campbell-al.-2004} to recover the parameters of the continuous-time system eqs~\eqref{eq:cont1}--\eqref{eq:cont2} from the restricted VAR(1) eqs~\eqref{eq:disc1}--\eqref{eq:disc2}. However, \citet{Campbell-al.-2004} use the risk premium as state variable; we prefer to use the Sharpe ratio. In matrix form, the discrete-time VAR eqs~\eqref{eq:disc1}--\eqref{eq:disc2} is
\begin{equation}
\begin{pmatrix}
𝛥 \ln P_{t+1} - r^f \\ z_{t+1}
\end{pmatrix}
=
\begin{pmatrix}
a_r \\ a_z
\end{pmatrix}
+
\begin{pmatrix}
0 & b_r \\ 0 & b_z
\end{pmatrix}
\begin{pmatrix}
𝛥 \ln P_t - r^f \\ z_t
\end{pmatrix}
+
\begin{pmatrix}
𝜀^r_{t+1} \\ 𝜀^z_{t+1}
\end{pmatrix}
\label{eq:disc-matrix}
\end{equation}

The first step is to aggregate the continuous-time model over a span of time taking point observations at evenly spaced points~$\{ t_0, t_1, …, t_n, t_{n+1}, … \}$, with~$𝛥t = t_n - t_{n-1}$. We then obtain, using the discretization method developed by \citet{Bergstrom-1984}
\begin{align}
\begin{pmatrix}
𝛥 \ln P╻ - r 𝛥t \\ X╻
\end{pmatrix}
= {}
&\begin{pmatrix}
(-𝜎^2/2 + 𝜎𝜃)𝛥t - (1-ⅇ^{-𝜅𝛥t}) \frac{𝜎𝜃}{𝜅} \\ (1-ⅇ^{-𝜅𝛥t})𝜃
\end{pmatrix}
+ {}\nonumber\\
&\begin{pmatrix}
1 & (1-ⅇ^{-𝜅𝛥t}) \frac{𝜎}{𝜅} \\ 0 & ⅇ^{-𝜅𝛥t}
\end{pmatrix}
\begin{pmatrix}
𝛥 \ln P_{t_n} - r \\ X_{t_n}
\end{pmatrix}
+
\begin{pmatrix}
U^p╻ \\ U^x╻
\end{pmatrix}
\label{eq:cont-matrix}
\end{align}
where
\begin{equation}
\begin{pmatrix}
U^p╻\\U^x╻
\end{pmatrix}
= \int_{𝜏=0}^{𝛥t}
\begin{pmatrix}
1 & (1-ⅇ^{-𝜅𝛥t}) \frac{𝜎}{𝜅} \\
0 & ⅇ^{-𝜅𝛥t}
\end{pmatrix}
\begin{pmatrix}
𝜎 & 0\\
𝜁𝜌 & 𝜁\sqrt{1{-}𝜌^2}
\end{pmatrix}
\begin{pmatrix}
ⅆB^p_{t_n+𝜏}\\ⅆZ^x_{t_n+𝜏}
\end{pmatrix} \label{eq:u_matrix}
\end{equation}
with~$ⅆB^x_t = 𝜌 ⅆB^p_t + \sqrt{1{-}𝜌^2} ⅆZ^x_t$ where~$B^p_t$ and~$Z^x_t$ are two independent brownian terms.

The second step is to apply a linear transformation for the process~$X_t$ in~\eqref{eq:cont-matrix} so that we can relate the parameters of the transformed system to the parameters of the matrix form~\eqref{eq:disc-matrix} of the discrete-time VAR model. Thus, when we normalize time span~$𝛥t = 1$, since everything is in quarter, we get (for~$b_z, b_r > 0$)
\begin{align}
r & = r^f \label{eq:r} \\
𝜃 & = \frac{a_z b_r}{𝜎_r(1{-}b_z)} + \frac{a_r + 𝜎_r^2/2}{𝜎_r} \label{eq:theta}\\
𝜅 & = - \ln(b_z) \label{eq:kappa} \\
𝜎 & = 𝜎_r \label{eq-sigma} \\
𝜁 & = b_r \frac{𝜎_z}{𝜎_r} \label{eq:Sigma_theta}\\
𝜌 & = \frac{𝜎_{rz}}{𝜎_r \, 𝜎_z} \label{eq-rho}
\end{align}
The appendix proves these results. Table~\ref{tab:Recovering} shows the value of the parameters of the continuous-time equivalent VAR implied by the \citet{Brandt-al.-2005} estimates.

\begin{table}
\centering
\begin{tabular}{c>{\qquad}c}
\toprule
\textbf{Discrete-time world} & \textbf{Continuous-time world} \tabularnewline
\midrule
\multicolumn{2}{c}{\textbf{Models}}\tabularnewline
\citet{Brandt-al.-2005} & \citet{Kim-Omberg-1996} \tabularnewline
\begin{tabular}{r<{~=~}@{}l}
$𝛥 \ln P_{t+1} - r^f$ & $a_r + b_r z_t + 𝜀^r_{t+1}$ \tabularnewline
$z_{t+1}$ & $a_z + b_z z_t + 𝜀^z_{t+1}$ \tabularnewline
$\mathrm{V}(𝜀^r_t)$ & $𝜎^2_r$ \tabularnewline
$\mathrm{V}(𝜀^z_t)$ & $𝜎^2_s$ \tabularnewline
$\mathrm{Cov}(𝜀^r_t, 𝜀^z_t)$ & $ 𝜎_{rz}$
\end{tabular}
&
\begin{tabular}{r<{~=~}@{}l}
$ⅆ P^f_t/P^f_t$ & $r ⅆt$ \tabularnewline
$ⅆ P_t/P_t$ & $(𝜎 X_t + r) ⅆ t + 𝜎 ⅆB_t^p$ \tabularnewline
$ⅆ X_t$ & $𝜅(𝜃 - X_t) ⅆt + 𝜁 ⅆB_t^x$ \tabularnewline
$ⅆB_t^p ⅆB_t^x$ & $𝜌 ⅆt$ \tabularnewline
\end{tabular} \tabularnewline
\midrule
\multicolumn{2}{c}{\textbf{Parameter values}} \tabularnewline
\citet{Brandt-al.-2005} & Our computations eqs~\eqref{eq:r}--\eqref{eq-rho} \tabularnewline
\begin{tabular}{cr}
$r^f$ & \num{.015} \tabularnewline
$a_r$ & \num{.227} \tabularnewline
$b_r$ & \num{.060} \tabularnewline
$a_z$ & \num{-.155} \tabularnewline
$b_z$ & \num{.958} \tabularnewline
$𝜎^2_r$ & \num{.0060} \tabularnewline
$𝜎^2_z$ & \num{.0049} \tabularnewline
$𝜎_{rz}$ & \num{-.0051} \tabularnewline
\end{tabular}
&
\begin{tabular}{cr}
$r$ & \num{.015} \tabularnewline
$𝜃$ & \num{0.111} \tabularnewline
$𝜅$ & \num{0.0429} \tabularnewline
$𝜎$ & \num{.0060} \tabularnewline
$𝜁$ & \num{0.0542} \tabularnewline
$𝜌$ & \num{-0.941} \tabularnewline
\end{tabular} \tabularnewline
\bottomrule
\end{tabular}
\caption{Recovering continuous-time parameters}
\label{tab:Recovering}
\end{table}

\section{Portfolio choice problem in continuous time with CRRA preferences}

We can now solve the portfolio choice problem of the investor with long-term horizons who faces to the investment opportunity set described in the previous section. We rely on the recent advances in~\citet{Honda-Kamimura-2011} who use the verification theorem and show that the explicit solution provided under continuous time is in fact an optimal solution especially for risk aversion greater that one.

We consider an investor with initial wealth~$W_{t_0} > 0$ who has only two assets (riskless short-term bond and stocks) available for investment. The financial markets are incomplete. Furthermore, the investor can undertake continuous trading, he has no labor income and only cares about terminal wealth~$W_T$ where~$T$ is the finite planning horizon. The dynamics of price changes are described by~\eqref{eq:Pf} and~\eqref{eq:cont1}--\eqref{eq:cont2}. If~$𝛼_t$ is the share of wealth invested in stocks, the instantaneous wealth would be given by
\begin{equation}
\frac{ⅆ W_t}{W_t} = 𝛼_t\frac{ⅆ P_t}{P_t} + (1{-}𝛼_t) \frac{ⅆ P^f_t}{P^f_t}
\end{equation}
Properly substituting the dynamics of~$ⅆ P_t/P_t$  and~$ⅆ P^f_t/P^f_t$, wealth dynamics (also called the budget constraint) becomes:
\begin{equation}
ⅆ W_t = (𝛼_t 𝜎 X_t + r)W_t ⅆ t + 𝛼_t 𝜎 W_t ⅆ B^p_t
\label{eq:budget-constraint}
\end{equation}
Notice that wealth process reflects uncertainty in instantaneous returns (term~$ⅆ B^p_t$) and about the state variable (the term~$X_t$). Given this formalization about wealth process, at time~$t_0$, the investor's optimization problem can then be expres\-sed as
\begin{equation}
\max_{𝛼_{t_0}} \quad \mathrm{E}_{t_0} \, ⅇ^{-𝛽 T} u(W_T) \quad \text{subject to the constraint \eqref{eq:budget-constraint}} \quad W_{t_0} \text{ fixed}
\label{eq:J}
\end{equation}
where~$\mathrm{E}_{t_0}$ denotes the operator of conditional rational expectation at date~$t_0$, $𝛽$ the time discount parameter (with~$𝛽  > 0$) and~$u(∙)$ the utility function defined over terminal wealth. Let~$J(W_{t_0}, X_{t_0}, t_0)$ defines the value of the problem defined in~\eqref{eq:J} at time~$t_0$
\begin{equation}
J(W_{t_0}, X_{t_0}, t_0) = \max_{𝛼_{t_0}} \quad \mathrm{E}_{t_0} \, ⅇ^{-𝛽 T} u(W_T)
\label{eq:J}
\end{equation}
The Bellman equation generalizes this problem for every time~$t$ so that
\begin{equation}
J(W_t, X_t, t) = \max_{𝛼_t} ~ \mathrm{E}_t \, J(W_t{+}ⅆ W_t, X_t{+}ⅆ X_t, t{+}ⅆ t)
\label{eq:Bellman_0}
\end{equation}
Equation~\eqref{eq:Bellman_0} emphasizes the fact that current optimal decisions depend on the conditional expected value of the problem which, in turn, is intimately linked to future wealth and the state variable. Applying Ito's lemma to the Bellman equation, we find that
\begin{align}
0 = \max_{𝛼_t} ~ \bigg[ \frac{∂J}{∂W_t} (𝛼_t 𝜎 X_t + r) W_t &{} + \frac{∂J}{∂t} + \frac{∂J}{∂X_t} 𝜅 (𝜃 - X_t) + {} \nonumber \\
& \frac{1}{2} \frac{∂^2 J}{∂^2 W_t} 𝜎^2 𝛼^2_t W^2_t + \frac{1}{2} \frac{∂^2 J}{∂^2 X_t} 𝜁^2 + \frac{∂^2J}{∂W_t ∂X_t} 𝜎 𝛼_t 𝜁 𝜌 W_t \bigg] \label{eq:Bellman-Taylor}
\end{align}

The first order condition of equation~\eqref{eq:Bellman-Taylor} with respect to~$𝛼_t$ implies that
\begin{equation}
𝛼^\star_t = \frac{∂J / ∂ W_t}{∂^2J / ∂^2W_t} \frac{1}{W_t} \frac{X_t}{𝜎} + \frac{∂^2 J / (∂W_t ∂X_t)}{∂^2J / ∂^2W_t} \frac{1}{W_t} \frac{𝜁}{𝜎} 𝜌
\label{Eq:alpha_op}
\end{equation}
\citet{Merton-1971} was the first to propose such additive decomposition between a \emph{myopic demand} (first term) and a \emph{hedging demand} (second term) of the optimal allocation to stocks. There is no hedging demand especially when the opportunity set is nonstochastic ($𝜁 = 0$) or when the opportunity set is uncorrelated with asset returns ($𝜌 = 0$).

Now, we need to explicitly define the function~$J(∙)$. The first conjecture (see \citet{Kim-Omberg-1996}) is to assume
\begin{equation}
J(W_t, X_t, t) = ⅇ^{-𝛽 t} u(W_t) \, [f(X_t, t)]^𝛾
\end{equation}
where~$f(∙)$ is an auxiliary function with the terminal condition~$f(X_T, T) = 1$. We consider the CRRA preferences~$u(W_t) = W_t^{1-𝛾}/(1{-}𝛾)$ where~$𝛾$ is the coefficient of relative risk aversion. Thus, the hedging demand in~\eqref{Eq:alpha_op} could straightforward be expressed as
\[
\frac{∂ f / ∂ X_t}{f} \frac{𝜁}{𝜎} 𝜌 = \frac{∂ \ln f}{∂ X_t} \frac{𝜁}{𝜎} 𝜌
\]
Then, under CRRA preferences hypothesis, the optimal allocation to stocks could be expressed as
\begin{equation}
𝛼^\star_t = \frac{1}{𝛾} \frac{X_t}{𝜎} + \frac{∂ \ln f}{∂ X_t} \frac{𝜁}{𝜎} 𝜌 \label{eq:alpha-star}
\end{equation}
So, the Bellman equation~\eqref{eq:Bellman-Taylor} can be rewritten as
\begin{align}
0 = &\frac{f'_t}{f} + \frac{1{-}𝛾}{𝛾} r - \frac{𝛽}{𝛾} + \frac{1{-}𝛾}{2\,𝛾^2} X_t^2 + \frac{f'_x}{f} \frac{1{-}𝛾}{𝛾} 𝜁 X_t 𝜌  + \frac{f'_x}{f} 𝜅 (𝜃 - X_t) + {} \nonumber \\
&\frac{f''_{xx}}{f} \frac{𝜁^2}{2} + \left( \frac{f'_x}{f} \right)^2 \frac{1{-}𝛾}{2} 𝜁^2 (𝜌^2-1) \label{eq:Bellman-Taylor-2}
\end{align}
where we use intuitive notations for the derivatives of the function~$f(∙)$. Equation~\eqref{eq:Bellman-Taylor-2} is a partial differential equation which admits analytical solutions especially if utility is logarithmic ($𝛾 = 1$ by l'Hopital's rule) or if markets are complete ($𝜌 = \pm 1$).

The second conjecture is to assume 
\begin{equation}
f(X_t, t) = \exp \left[ C_0(t) + C_1(t) \, X_t + \frac{1}{2} C_2(t) \, X^2_t \right]
\end{equation}
where~$C_0(t)$, $C_1(t)$ and~$C_2(t)$ are some undetermined time varying coefficients (with~$C_0(T) = C_1(T) = C_2(T) = 0$). Under this conjecture, using equation~\eqref{eq:alpha-star}, the optimal allocation to stocks is
\begin{equation}
𝛼^\star_t = \frac{1}{𝛾} \frac{X_t}{𝜎} + [C_1(t) + C_2(t) \, X_t] \frac{𝜁}{𝜎} 𝜌
\label{eq:alpha_C}
\end{equation}
We only need to recover~$C_1(t)$ and~$C_2(t)$ coefficients.

This conjecture was also used by~\citet{Kim-Omberg-1996} and by~\citet{Liu-2007} among others. More recently, \citet{Honda-Kamimura-2011} show that the explicit solution derived from the Bellman equation is in fact, even if the markets are incomplete, an optimal solution to the problem of the long-term investor who only care about terminal wealth and who have a risk aversion larger than unity.

Let us substitute our second conjecture in the equation~\eqref{eq:Bellman-Taylor-2}
\begin{align}
0 = \left[ \frac{ⅆ C_2}{ⅆ t} + a\,C_2^2 + b\,C_2 + c \right] X^2_t & + {}
\left[ \frac{ⅆ C_1}{ⅆ t} + \frac{b}{2} C_1 + 𝜅 𝜃 C_2 + a \, C_1 C_2 \right] X_t + {} \nonumber \\
& \left[ \frac{ⅆ C_0}{ⅆ t} + \frac{1{-}𝛾}{𝛾} r - \frac{𝛽}{𝛾} + 𝜅 𝜃 C_1 + \frac{𝜁^2}{2} C_2 + \frac{a}{2} C_1^2 \right]
\label{eq:system_0}
\end{align}
where~$a = [1 + (1{-}𝛾)(𝜌^2{-}1)] \, 𝜁^2$, $b = 2[(1{-}𝛾)/𝛾 𝜁 \, 𝜌 - 𝜅]$ and~$c = (1{-}𝛾)/𝛾^2$. As, whatever the value of~$X_t$, the equation~\eqref{eq:system_0} must hold, all terms within brackets are simultaneously set to zero to solve the equation for~$C_0(∙)$, $C_1(∙)$, and~$C_2(∙)$. Defining the discriminant~$D$
\[
D = b^2 - 4\,a\,c
\]
one can check that if~$𝛾 > 1$ then~$D > 0$. Thus, the two solutions of interest are given by
\begin{align}
\label{Eq:C_2}
C_2(t) & = \frac{2\,c \left(1-ⅇ^{-𝛿(T-t)}\right)} {2𝛿-\left(b+𝛿\right)\left(1-ⅇ^{-𝛿(T-t)}\right)} \\
\label{Eq:C_1}
C_1(t) & = \frac{4\,c\,𝜅𝜃}{𝛿}\,\frac{\left(1-ⅇ^{-𝛿(T-t)/2}\right)^2}{2𝛿-\left(b+𝛿\right)\left(1-ⅇ^{-𝛿(T-t)}\right)}
\end{align}
where~$𝛿$ denotes~$\sqrt D$. \citet{Kim-Omberg-1996} called this the normal solution and discussed about some alternative solutions those are beyond the scope of this paper. The appendix provides details about~\eqref{Eq:C_2} and~\eqref{Eq:C_1}.  It is easy to see that there is a linear relation between~$C_1(∙)$ and~$C_2(∙)$. Then, for~$𝛾 > 1$, $C_1$ and~$C_2$ are always negative. As a result, since~$𝜌 < 0$, the hedging demand is always positive when the preferences are not logarithmic (more precisely for~$𝛾 > 1$) and the market price of risk is positive. 
\section{Numerical results}

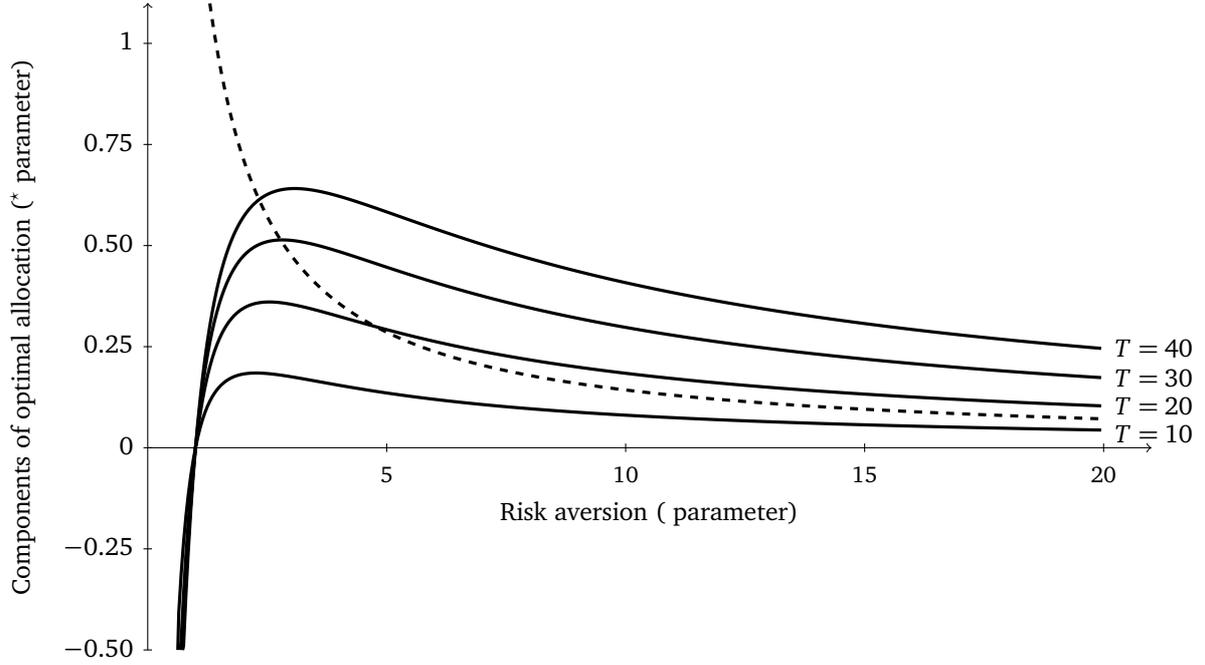
\begin{figure}
	\centering
	\def\x{0}\def\xx{21}\def\xsecond{10}\def\xrounddigits{0}
	\def\y{-.5}\def\yy{1.1}\def\ysecond{-.25}\def\yrounddigits{2}
	\def\xlabel{Risk aversion ($𝛾$ parameter)}
	\def\ylabel{Components of optimal allocation ($𝛼^\star$ parameter)}
	\def\xwidth{.8\textwidth}\def\ratioxy{.65}\def\grad{.75}
	\pgfmathsetmacro{\xscale}{\xwidth/1cm/(\xx-\x)}
	\pgfmathsetmacro{\yscale}{\xscale*(\xx-\x)/(\yy-\y)*\ratioxy}
	\pgfmathsetmacro{\xgrad}{.75mm*\grad/(\xwidth*\ratioxy)*(\yy-\y)}
	\pgfmathsetmacro{\ygrad}{.75mm*\grad/\xwidth*(\xx-\x)}
	\pgfmathsetmacro{\xfirst}{5}
	\pgfmathsetmacro{\xlast}{.99*\xx}
	\pgfmathsetmacro{\ylast}{.95*\yy}
	\begin{tikzpicture}[xscale=\xscale,yscale=\yscale]
		\draw [->] (\x, 0) -- (\xx, 0) ;                 
		\foreach \ix in {\xfirst, \xsecond, ..., \xlast} 
			\draw (\ix, 0+\xgrad) -- (\ix, 0-\xgrad) node[anchor=north]
				{\pgfmathparse{\ix}\footnotesize\strut\num[group-digits = true, round-mode = places, round-precision = \xrounddigits]{\pgfmathresult}} ;
		\draw (.5*\x+.5*\xx, 0) node[anchor=north] 
			{\shortstack{\strut\\\small \xlabel}} ;
		\draw [->] (\x, \y) -- (\x, \yy) ;          
		\foreach \iy in {\y, \ysecond, ..., \ylast} 
			\draw (\x+\ygrad, \iy) -- (\x-\ygrad, \iy) node[anchor=east]
				{\pgfmathparse{\iy}\small\strut\num[group-digits = true, round-mode = places, round-precision = \yrounddigits]{\pgfmathresult}} ;
		\draw (\x, .5*\y+.5*\yy) node[anchor=east]
			{\rotatebox{90}{\raisebox{8ex}{\small \ylabel}}} ;
	\draw [very thick, black] plot [] file{explicit-solution-hd-wrt-gamma-10.txt} ;
	\draw (20, 0.04383) node[anchor=west] {\small \raisebox{-2ex}{$T=10$}} ;
	\draw [very thick, black] plot [] file{explicit-solution-hd-wrt-gamma-20.txt} ;
	\draw (20, 0.10366) node[anchor=west] {\small $T=20$} ;
	\draw [very thick, black] plot [] file{explicit-solution-hd-wrt-gamma-30.txt} ;
	\draw (20, 0.17345) node[anchor=west] {\small $T=30$} ;
	\draw [very thick, black] plot [] file{explicit-solution-hd-wrt-gamma-40.txt} ;
	\draw (20, 0.24573) node[anchor=west] {\small $T=40$} ;
	\draw [very thick, black, dashed] plot [] file{explicit-solution-md-wrt-gamma.txt} ;
	\end{tikzpicture}
	\caption{Myopic (dashed line) and hedging (solid line) demands as function of risk aversion for $X_{t_0} = 𝜃$}
	\label{fig:Myopic-hedging}
\end{figure}
\begin{table}
	\centering \small \tabcolsep1\tabcolsep
	\newbox{\abox}\savebox{\abox}{%
	\begin{tabular}{>{\bfseries} c >{\bfseries} c *5r @{\quad\quad} *5r}
		\toprule
		& & \multicolumn{5}{l}{$𝛾=5$} & \multicolumn{5}{l}{$𝛾=15$} \tabularnewline
		\cmidrule(r){3-7}\cmidrule(l){8-12}
		$T$ &&
		\multicolumn{1}{c}{$X_{(10)}$}&
		\multicolumn{1}{c}{$X_{(30)}$}&
		\multicolumn{1}{c}{$X_{(50)}$}&
		\multicolumn{1}{c}{$X_{(70)}$}&
		\multicolumn{1}{c}{$X_{(90)}$}&
		\multicolumn{1}{c}{$X_{(10)}$}&
		\multicolumn{1}{c}{$X_{(30)}$}&
		\multicolumn{1}{c}{$X_{(50)}$}&
		\multicolumn{1}{c}{$X_{(70)}$}&
		\multicolumn{1}{c}{$X_{(90)}$}\tabularnewline
		\midrule
10 & MD & \num{-34.0} & \num{3.0} & \num{28.6} & \num{54.2} & \num{91.1} & \num{-11.3} & \num{1.0} & \num{9.5} & \num{18.1} & \num{30.4} \tabularnewline
 & HD & \num{-10.9} & \num{3.5} & \num{13.5} & \num{23.6} & \num{38.0} & \num{-4.6} & \num{1.5} & \num{5.7} & \num{9.9} & \num{15.9} \tabularnewline
\midrule
20 & MD & \num{-34.0} & \num{3.0} & \num{28.6} & \num{54.2} & \num{91.1} & \num{-11.3} & \num{1.0} & \num{9.5} & \num{18.1} & \num{30.4} \tabularnewline
 & HD & \num{-15.9} & \num{10.8} & \num{29.2} & \num{47.7} & \num{74.3} & \num{-7.2} & \num{4.9} & \num{13.3} & \num{21.6} & \num{33.7} \tabularnewline
\midrule
30 & MD & \num{-34.0} & \num{3.0} & \num{28.6} & \num{54.2} & \num{91.1} & \num{-11.3} & \num{1.0} & \num{9.5} & \num{18.1} & \num{30.4} \tabularnewline
 & HD & \num{-16.0} & \num{19.8} & \num{44.7} & \num{69.5} & \num{105.3} & \num{-7.7} & \num{9.8} & \num{21.9} & \num{34.1} & \num{51.6} \tabularnewline
\midrule
40 & MD & \num{-34.0} & \num{3.0} & \num{28.6} & \num{54.2} & \num{91.1} & \num{-11.3} & \num{1.0} & \num{9.5} & \num{18.1} & \num{30.4} \tabularnewline
 & HD & \num{-13.2} & \num{29.1} & \num{58.3} & \num{87.6} & \num{129.8} & \num{-6.5} & \num{15.5} & \num{30.7} & \num{46.0} & \num{68.0} \tabularnewline
		\bottomrule
		\end{tabular}%
	}
	\centering
	\begin{minipage}{\widthof{\usebox{\abox}}-\widthof{~~}}
		{\centering \makebox[0pt]{\usebox{\abox}}\par }
		\footnotesize For each risk aversion~$𝛾$, the first line reports the myopic demand (\textbf{MD}) and the second line the hedging demand (\textbf{HD}) without short selling constraints. We present the results for 5 different initial values of the Sharpe ratio~$X$. Each value corresponds to the~$p$-th percentile of the unconditional distribution of~$X$, defined by equations~\eqref{eq:Unc-dist} and denoted by~$X_{(p)}$, where~$p$ takes values 10, 30, 50, 70, and 90 (then~$X_{(50)} = 𝜃$).
	\end{minipage}
	\caption{Myopic and hedging demands for investment horizon of $T$ quarters (\%)}
	\label{tab:Myopic-hedging}
\end{table}

As mentioned above, we illustrate our approach using the well documented \citet{Brandt-al.-2005} example. Table~\ref{tab:Recovering} collects the continuous-time parameters recovered from this example. For comparison purposes, we also use the \citet{Garlappi-Skoulakis-2009} results, obtained from the same discrete-time VAR(1) estimates and by means of a sophisticated numerical method.

Figure~\ref{fig:Myopic-hedging} and table~\ref{tab:Myopic-hedging} help to understand the long-term investor problem. For~$𝛾 = 1$ \emph{i.e.} the case of logarithmic utility, no hedging demand is required. For this case, the dynamic portfolio choice reduces to static one whatever the time horizon. Otherwise, for~$ 𝛾 > 1$ and horizon longer than one, under CRRA preferences and mean reverting returns, agent should have a positive hedging demand to prevent adverse changes in investment opportunities \citep{Merton-1971}.  However, for~$𝛾 \to \infty$, more specifically for a very conservative agent, stocks are not attractive. Thus, he would not invest into stocks since the total demand (sum of myopic demand and hedging demand) converges toward zero. Our results reset all theses basic important features.

The total demand is sensitive to risk aversion. Results from previous studies imply that myopic and hedging demands are more sensitive to small values of risk aversion. We confirm this and argue that the sensitivity of hedging demand to state variable is maximal near the critical point~$𝛾 \approx 2$. Our equation~\eqref{eq:alpha_C} and figure~\ref{fig:Myopic-hedging} show this evidence. To quantitatively see this, just evaluate the derivative of~$𝛼^\star$ with respect to~$X$. 

Table~\ref{tab:Myopic-hedging} reports the evidence that both myopic and hedging demand are sensitive to initial value of Sharpe ratio. These two components of optimal allocation individually increase with the percentile of the Sharpe ratio unconditional distribution. Thus, the total demand exhibits the same behavior. This is consistent with \citet{Campbell-al.-2004} among others. In fact, high Sharpe ratio or equivalently high risk premium relative to volatility signals better investment opportunities. Therefore, optimal fraction to allocate into stocks should increase from the knowledge of mean reverting parameter that serves to quantify the expected Sharpe ratio. 

Myopic demand is independent from time horizon while hedging demand increases nonlinearly with time horizon. However, table~\ref{tab:Myopic-hedging} quantitative figures suggest that this relation is almost linear but small changes in horizon induce substantial hedging demand. Horizon effect is important but quiet monotonic for a given percentile of the state variable unconditional distribution. All changes in total demand for fixed risk aversion and state variable are due to changes in horizon and are large for small risk aversion.

The horizon effect on hedging demand is important in optimal allocation because it widely dominates for longer horizons. In fact, when horizon is greater than 20 quarters, hedging demand becomes always greater than myopic demand when the Sharpe ratio initial value is between~30 and~70 percentiles.

\begin{table}
	\centering \small \tabcolsep1\tabcolsep
	\newbox{\abox}\savebox{\abox}{%
	\begin{tabular}{>{\bfseries} c >{\bfseries} c *5r @{\quad\quad} *5r}
		\toprule
		& & \multicolumn{5}{l}{$𝛾=5$} & \multicolumn{5}{l}{$𝛾=15$} \tabularnewline
		\cmidrule(r){3-7}\cmidrule(l){8-12}
		$T$ &&
		\multicolumn{1}{c}{$X_{(10)}$}&
		\multicolumn{1}{c}{$X_{(30)}$}&
		\multicolumn{1}{c}{$X_{(50)}$}&
		\multicolumn{1}{c}{$X_{(70)}$}&
		\multicolumn{1}{c}{$X_{(90)}$}&
		\multicolumn{1}{c}{$X_{(10)}$}&
		\multicolumn{1}{c}{$X_{(30)}$}&
		\multicolumn{1}{c}{$X_{(50)}$}&
		\multicolumn{1}{c}{$X_{(70)}$}&
		\multicolumn{1}{c}{$X_{(90)}$}\tabularnewline
		\midrule
10 & LT & \num{0.0} & \num{6.5} & \num{42.1} & \num{77.7} & \num{100.0} & \num{0.0} & \num{2.5} & \num{15.2} & \num{27.9} & \num{46.3} \tabularnewline
 & GS & \num{0.0} & \num{13.3} & \num{43.2} & \num{73.1} & \num{100.0} & \num{0.0} & \num{4.3} & \num{15.4} & \num{27.0} & \num{44.7} \tabularnewline
 & $𝛥$ & \num{0.0} & \num{-6.8} & \num{-1.1} & \num{4.6} & \num{0.0} & \num{0.0} & \num{-1.8} & \num{-0.2} & \num{0.9} & \num{1.6} \tabularnewline
\midrule
20 & LT & \num{0.0} & \num{13.7} & \num{57.8} & \num{100.0} & \num{100.0} & \num{0.0} & \num{5.9} & \num{22.8} & \num{39.7} & \num{64.1} \tabularnewline
 & GS & \num{0.0} & \num{24.4} & \num{57.2} & \num{89.7} & \num{100.0} & \num{0.0} & \num{10.7} & \num{25.1} & \num{40.4} & \num{63.2} \tabularnewline
 & $𝛥$ & \num{0.0} & \num{-10.7} & \num{0.6} & \num{10.3} & \num{0.0} & \num{0.0} & \num{-4.8} & \num{-2.3} & \num{-0.7} & \num{0.9} \tabularnewline
\midrule
30 & LT & \num{0.0} & \num{22.8} & \num{73.2} & \num{100.0} & \num{100.0} & \num{0.0} & \num{10.8} & \num{31.5} & \num{52.1} & \num{81.9} \tabularnewline
 & GS & \num{0.0} & \num{32.8} & \num{68.4} & \num{100.0} & \num{100.0} & \num{0.0} & \num{17.5} & \num{35.2} & \num{54.0} & \num{80.7} \tabularnewline
 & $𝛥$ & \num{0.0} & \num{-10.0} & \num{4.8} & \num{0.0} & \num{0.0} & \num{0.0} & \num{-6.7} & \num{-3.7} & \num{-1.9} & \num{1.2} \tabularnewline
\midrule
40 & LT & \num{0.0} & \num{32.0} & \num{86.9} & \num{100.0} & \num{100.0} & \num{0.0} & \num{16.5} & \num{40.2} & \num{64.0} & \num{98.3} \tabularnewline
 & GS & \num{0.0} & \num{38.8} & \num{77.6} & \num{100.0} & \num{100.0} & \num{0.0} & \num{24.1} & \num{44.5} & \num{65.7} & \num{94.6} \tabularnewline
 & $𝛥$ & \num{0.0} & \num{-6.8} & \num{9.3} & \num{0.0} & \num{0.0} & \num{0.0} & \num{-7.6} & \num{-4.3} & \num{-1.7} & \num{3.7} \tabularnewline
		\bottomrule
		\end{tabular}%
	}
	\centering
	\begin{minipage}{\widthof{\usebox{\abox}}-\widthof{~~}}
		{\centering \makebox[0pt]{\usebox{\abox}}\par }
		\footnotesize For each risk aversion~$𝛾$, the first line reports our results (\textbf{LT} – optimal allocation to stocks in continuous time), the second line the \citet{Garlappi-Skoulakis-2009} results (\textbf{GS} – optimal allocation to stocks in discrete time), and the third line reports the difference between our results and \citet{Garlappi-Skoulakis-2009} results. We present the results for 5 different initial values of the Sharpe ratio~$X$ calibrated using the same estimates involving dividend price ratio as in \textbf{GS}. Each value corresponds to the~$p$-th percentile of the unconditional distribution of~$X$, defined by equation~\eqref{eq:Unc-dist} and denoted by~$X_{(p)}$, where~$p$ takes values 10, 30, 50, 70, and 90.
\end{minipage}
	\caption{Optimal allocation to stocks for investment horizon of $T$ quarters (\%)}
	\label{tab:Optimal}
\end{table}

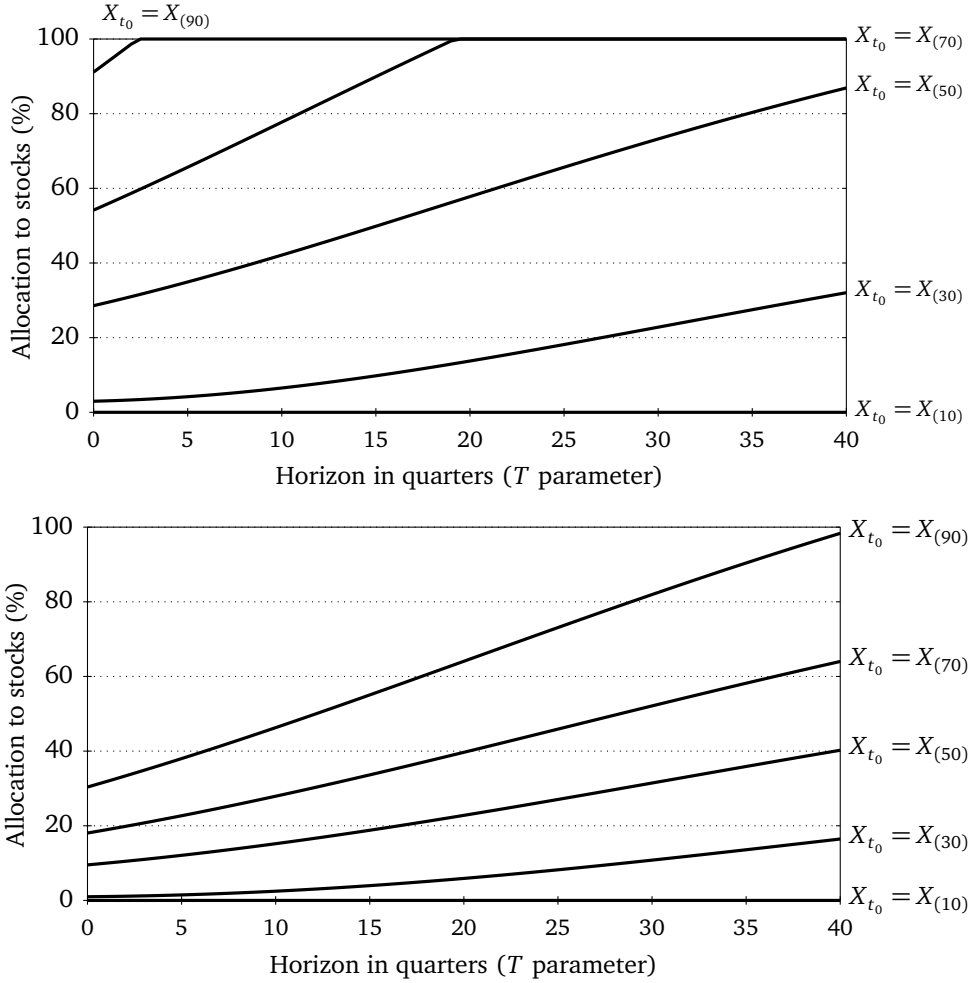
\begin{figure}
	\centering
	\def\x{0}\def\xx{40}\def\xsecond{5}\def\xrounddigits{0}
	\def\y{0}\def\yy{100}\def\ysecond{20}\def\yrounddigits{0}
	\def\xlabel{Horizon in quarters ($T$ parameter)}
	\def\ylabel{Allocation to stocks (\%)}
	\def\xwidth{.6\textwidth}\def\ratioxy{.5}\def\grad{.75}
	\pgfmathsetmacro{\xscale}{\xwidth/1cm/(\xx-\x)}
	\pgfmathsetmacro{\yscale}{\xscale*(\xx-\x)/(\yy-\y)*\ratioxy}
	\pgfmathsetmacro{\xgrad}{.75mm*\grad/(\xwidth*\ratioxy)*(\yy-\y)}
	\pgfmathsetmacro{\ygrad}{.75mm*\grad/\xwidth*(\xx-\x)}
	\begin{tikzpicture}[xscale=\xscale,yscale=\yscale]
		\draw (\x, \y) -- (\xx, \y) ;            
		\draw (\x, \yy) -- (\xx, \yy) ;          
		\foreach \ix in {\x, \xsecond, ..., \xx} 
			\draw (\ix, \y+\xgrad) -- (\ix, \y-\xgrad) node[anchor=north]
				{\pgfmathparse{\ix}\footnotesize\strut\num[group-digits = true, round-mode = places, round-precision = \xrounddigits]{\pgfmathresult}} ;
		\draw (.5*\x+.5*\xx, 0) node[anchor=north] 
			{\shortstack{\strut\\\small \xlabel}} ;
		\draw (\x, \y) -- (\x, \yy) ;            
		\draw (\xx, \y) -- (\xx, \yy) ;          
		\foreach \iy in {\y, \ysecond, ..., \yy} 
			\draw (\x+\ygrad, \iy) -- (\x-\ygrad, \iy) node[anchor=east]
				{\pgfmathparse{\iy}\small\strut\num[group-digits = true, round-mode = places, round-precision = \yrounddigits]{\pgfmathresult}} ;
		\foreach \iy in {\y, \ysecond, ..., \yy} 
			\draw (\xx, \iy) -- (\xx-\ygrad, \iy) ;
		\foreach \iy in {\y, \ysecond, ..., \yy} 
			\draw [dotted] (\x, \iy) -- (\xx-\ygrad, \iy) ;
		\draw (\x, .5*\y+.5*\yy) node[anchor=east]
			{\rotatebox{90}{\raisebox{4ex}{\small \ylabel}}} ;
	\draw [very thick, black] plot [] file{explicit-solution-wrt-T-5-0.100000.txt} ;
	\draw (40, 0) node[anchor=west] {\footnotesize $X_{t_0} = X_{(10)}$} ;
	\draw [very thick, black] plot [] file{explicit-solution-wrt-T-5-0.300000.txt} ;
	\draw (40, 32.046) node[anchor=west] {\footnotesize $X_{t_0} = X_{(30)}$} ;
	\draw [very thick, black] plot [] file{explicit-solution-wrt-T-5-0.500000.txt} ;
	\draw (40, 86.893) node[anchor=west] {\footnotesize $X_{t_0} = X_{(50)}$} ;
	\draw [very thick, black] plot [] file{explicit-solution-wrt-T-5-0.700000.txt} ;
	\draw (40, 100) node[anchor=west] {\footnotesize $X_{t_0} = X_{(70)}$} ;
	\draw [very thick, black] plot [] file{explicit-solution-wrt-T-5-0.900000.txt} ;
	\draw (0, 100) node[anchor=south west] {\footnotesize $X_{t_0} = X_{(90)}$} ;
	\end{tikzpicture}
	\def\x{0}\def\xx{40}\def\xsecond{5}\def\xrounddigits{0}
	\def\y{0}\def\yy{100}\def\ysecond{20}\def\yrounddigits{0}
	\def\xlabel{Horizon in quarters ($T$ parameter)}
	\def\ylabel{Allocation to stocks (\%)}
	\def\xwidth{.6\textwidth}\def\ratioxy{.5}\def\grad{.75}
	\pgfmathsetmacro{\xscale}{\xwidth/1cm/(\xx-\x)}
	\pgfmathsetmacro{\yscale}{\xscale*(\xx-\x)/(\yy-\y)*\ratioxy}
	\pgfmathsetmacro{\xgrad}{.75mm*\grad/(\xwidth*\ratioxy)*(\yy-\y)}
	\pgfmathsetmacro{\ygrad}{.75mm*\grad/\xwidth*(\xx-\x)}
	\begin{tikzpicture}[xscale=\xscale,yscale=\yscale]
		\draw (\x, \y) -- (\xx, \y) ;            
		\draw (\x, \yy) -- (\xx, \yy) ;          
		\foreach \ix in {\x, \xsecond, ..., \xx} 
			\draw (\ix, \y+\xgrad) -- (\ix, \y-\xgrad) node[anchor=north]
				{\pgfmathparse{\ix}\footnotesize\strut\num[group-digits = true, round-mode = places, round-precision = \xrounddigits]{\pgfmathresult}} ;
		\draw (.5*\x+.5*\xx, 0) node[anchor=north] 
			{\shortstack{\strut\\\small \xlabel}} ;
		\draw (\x, \y) -- (\x, \yy) ;            
		\draw (\xx, \y) -- (\xx, \yy) ;          
		\foreach \iy in {\y, \ysecond, ..., \yy} 
			\draw (\x+\ygrad, \iy) -- (\x-\ygrad, \iy) node[anchor=east]
				{\pgfmathparse{\iy}\small\strut\num[group-digits = true, round-mode = places, round-precision = \yrounddigits]{\pgfmathresult}} ;
		\draw (\x, .5*\y+.5*\yy) node[anchor=east]
			{\rotatebox{90}{\raisebox{4ex}{\small \ylabel}}} ;
		\foreach \iy in {\y, \ysecond, ..., \yy} 
			\draw (\xx, \iy) -- (\xx-\ygrad, \iy) ;
		\foreach \iy in {\y, \ysecond, ..., \yy} 
			\draw [dotted] (\x, \iy) -- (\xx-\ygrad, \iy) ;
	\draw [very thick, black] plot [] file{explicit-solution-wrt-T-15-0.100000.txt} ;
	\draw (40, 0) node[anchor=west] {\small $X_{t_0} = X_{(10)}$} ;
	\draw [very thick, black] plot [] file{explicit-solution-wrt-T-15-0.300000.txt} ;
	\draw (40, 16.456) node[anchor=west] {\small $X_{t_0} = X_{(30)}$} ;
	\draw [very thick, black] plot [] file{explicit-solution-wrt-T-15-0.500000.txt} ;
	\draw (40, 40.234) node[anchor=west] {\small $X_{t_0} = X_{(50)}$} ;
	\draw [very thick, black] plot [] file{explicit-solution-wrt-T-15-0.700000.txt} ;
	\draw (40, 64.012) node[anchor=west] {\small $X_{t_0} = X_{(70)}$} ;
	\draw [very thick, black] plot [] file{explicit-solution-wrt-T-15-0.900000.txt} ;
	\draw (40, 98.343) node[anchor=west] {\small $X_{t_0} = X_{(90)}$} ;
	\end{tikzpicture}
	\caption{Optimal allocation to stocks as function of the horizon for~$𝛾=5$ (first panel) and for~$𝛾=15$ (second panel) for 5 different initial values of the Sharpe ratio~$X$ (as in table~\ref{tab:Myopic-hedging} or~\ref{tab:Optimal})}
	\label{fig:Optimal}
\end{figure}

We finally use the common assumption of no-borrowing and short-sale constraints. Thus, in table~\ref{tab:Optimal}, we restrict all portfolio weights between~0 and~1.  One can notice that we generally obtain values fairly close to those of~\citet{Garlappi-Skoulakis-2009} while frameworks are not the same. \citet{Garlappi-Skoulakis-2009} worked in discrete-time and initial values of their state variable are drawn for the unconditional distribution of quarterly dividend price ratio. They use a sophisticated numerical optimization technique. We work in continuous time (no numerical optimization) and our initial values are computed using the unconditional distribution of continuous Sharpe ratio that we discretized in points observation and recovered using the same quarterly dividend price ratio.  However, a closer inspection of table~\ref{tab:Optimal} figures show that the optimal allocation to stocks is more sensitive to the state variable and to the time horizon than the sensitivity obtained by~\citet{Garlappi-Skoulakis-2009}. We run some numerical simulations, within the discrete-time framework, to evaluate our results in order to find the causes of the discrepancies between the two frameworks. We were unable to qualitatively and quantitatively invalidate our results. 

\begin{figure}
	\centering
	\def\x{0}\def\xx{10}\def\xsecond{1}\def\xrounddigits{0}
	\def\y{0}\def\yy{.25}\def\ysecond{.05}\def\yrounddigits{2}
	\def\xlabel{Quarters}
	\def\ylabel{Allocation to stocks}
	\def\xwidth{.6\textwidth}\def\ratioxy{.5}\def\grad{.75}
	\pgfmathsetmacro{\xscale}{\xwidth/1cm/(\xx-\x)}
	\pgfmathsetmacro{\yscale}{\xscale*(\xx-\x)/(\yy-\y)*\ratioxy}
	\pgfmathsetmacro{\xgrad}{.75mm*\grad/(\xwidth*\ratioxy)*(\yy-\y)}
	\pgfmathsetmacro{\ygrad}{.75mm*\grad/\xwidth*(\xx-\x)}
	\pgfmathsetmacro{\yyy}{1.0001*\yy}
	\begin{tikzpicture}[xscale=\xscale,yscale=\yscale]
		\draw (\x, \y) -- (\xx, \y) ;            
		\draw (\x, \yy) -- (\xx, \yy) ;          
		\foreach \ix in {\x, \xsecond, ..., \xx} 
			\draw (\ix, \y+\xgrad) -- (\ix, \y-\xgrad) node[anchor=north]
				{\pgfmathparse{\ix}\footnotesize\strut\num[group-digits = true, round-mode = places, round-precision = \xrounddigits]{\pgfmathresult}} ;
		\foreach \ix in {\x, \xsecond, ..., \xx} 
			\draw [dotted] (\ix, \y) -- (\ix, \yy) ;
		\draw (.5*\x+.5*\xx, 0) node[anchor=north] 
			{\shortstack{\strut\\\small \xlabel}} ;
		\draw (\x, \y) -- (\x, \yy) ;            
		\draw (\xx, \y) -- (\xx, \yy) ;          
		\foreach \iy in {\y, \ysecond, ..., \yyy} 
			\draw (\x+\ygrad, \iy) -- (\x-\ygrad, \iy) node[anchor=east]
				{\pgfmathparse{\iy}\small\strut\num[group-digits = true, round-mode = places, round-precision = \yrounddigits]{\pgfmathresult}} ;
		\foreach \iy in {\y, \ysecond, ..., \yy} 
			\draw (\xx, \iy) -- (\xx-\ygrad, \iy) ;
		\foreach \iy in {\y, \ysecond, ..., \yy} 
			\draw [dotted] (\x, \iy) -- (\xx-\ygrad, \iy) ;
		\draw (\x, .5*\y+.5*\yy) node[anchor=east]
			{\rotatebox{90}{\raisebox{7ex}{\small \ylabel}}} ;
	\draw [very thick, black, dashed] (0, .05) -- (1, .05) -- (2, .05) -- (2, .10) -- (3, .10) -- (4, .10) -- (5, .10) -- (6, .10) -- (7, .10) -- (8, .10)-- (8, .15) -- (9, .15) -- (10, .15) ;
	\draw [very thick, black] (0, 0.065) -- (1, 0.074) -- (2, 0.082) -- (3, 0.089) -- (4, 0.096) -- (5, 0.101) -- (6, 0.106) -- (7, 0.110) -- (8, 0.114) -- (9, 0.117) -- (10, 0.119) ;
	\end{tikzpicture}
	\caption{Path of optimal allocation to stocks for~$𝛾=5$, $X_0 = X_{(30)}$, and~$T = 10$ obtained by explicit solution (solid line) and by simulation and trial~$\{ \num{.05}, \num{.10}, \num{.15}, \num{.20}, \num{.25} \}$ grid (dashed line)}
	\label{fig:simulation}
\end{figure}
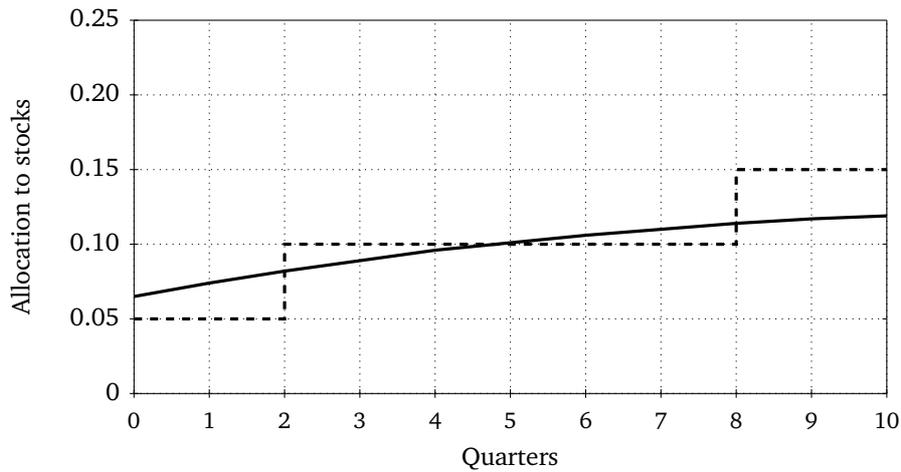
To test our results, we run some forward pure simulations in discrete time. More precisely, for instance, we explore the case where the initial value of the Sharpe ratio is the 30-th percentile ($X_0 = X_{(30)}$, the relative risk aversion is equal to~$5$ ($𝛾 = 5$), and the planning horizon is equal to~$10$ ($T = 10$ quarters). With this configuration, when we get an initial optimal allocation to stocks of~$\num{.065}$, \citet{Garlappi-Skoulakis-2009} obtain twice as many ($\num{.133}$, see table~\ref{tab:Optimal}). That's large. Thus, we first build a sample of size \num{100000} for~$z_{t+1}$, $z_{t+2}$, …, and~$z_{t+10}$ and for~$𝛥 \ln P_{t+1}$, $𝛥 \ln P_{t+2}$, …, and~$𝛥 \ln P_{t+10}$ using the restricted VAR(1) eqs~\eqref{eq:disc1}--\eqref{eq:disc2}. We choose the grid~$G = \{ \num{.05}, \num{.10}, \num{.15}, \num{.20}, \num{.25} \}$ for trial allocations to stocks, to overlay both our and \citet{Garlappi-Skoulakis-2009} solutions. Then, for each path in the sample, the value of terminal wealth is computed from the cartesian product~$G \times G \times ⋯ \times G$ of all possible strategies. The computational burden is very high as we evaluate~$5^{10} = \num{9765625}$ strategies. Figure~\ref{fig:simulation} shows that the forward path in discrete time (no numerical optimization) closes to the path of our explicit solution particularly at the critical starting point, the 30-th percentile of the state variable for small risk aversion ($𝛾 = 5$).

\section*{Conclusion}

We examine the ``continuous-time detour'' to solve the long-term investor problem when the stock returns are predictable. We obtain an explicit optimal solution in the continuous-time world and, after recovering the continuous-time parameters from the discrete-time world estimates, we reuse such solution to assess the sensitivities of optimal allocation to the initial values of the state variable, to the risk aversion and to the time horizon. We find greater sensitivities than those reported in the literature. We also find that the sensitivity of total demand to the state variable is not uniform along the unconditional distribution of the state variable.

Previous numerical approximation techniques that deal with the problem we consider are subject to some numerical errors. Therefore, they do not always provide accurate results. We show that the hedging demand part of allocation dominates at longer horizons and it is very sensitive to state variable especially when risk aversion decreases and/or the time horizon increases. This finding could explain the low accuracy of discrete numerical methods especially along the tails of the unconditional distribution of the state variable.    

\bibliography{Explicit-solution}

\section*{Appendix}

\subsection*{A.1 Proof of continuous VAR recovering by discrete VAR}

The matrix~\eqref{eq:disc-matrix} could be rewritten as 
\begin{align}
𝛥\ln P_{t+𝛥t} - r^{f} & =  a_r+b_rz_t+𝜀_{t+𝛥t}^{r} \label{eq:return_econ}\\
z_{t+𝛥t} & =  a_z+b_zz_t+𝜀_{t+𝛥t}^z \label{eq:dp_econ}
\end{align}
Equations~\eqref{eq:return_econ} and~\eqref{eq:dp_econ} describe a joint process of an econometric model in which~$z$ denotes the dividend price ratio. The corresponding discretized version of the continuous time model in matrix~\eqref{eq:cont-matrix} could be rewritten as
\begin{align}
𝛥\log P╻-r𝛥t & =  (-𝜎^2/2 + 𝜎𝜃)𝛥t - (1-ⅇ^{-𝜅𝛥t}) \frac{𝜎𝜃}{𝜅} + (1-ⅇ^{-𝜅𝛥t}) \frac{𝜎}{𝜅} X_{t_n} + U^p╻ \label{eq:return_cont}\\
X╻ & =  (1-ⅇ^{-𝜅𝛥t})𝜃+ⅇ^{-𝜅𝛥t} X_{t_n}+U^x╻ \label{eq:X_cont}
\end{align}
Comparing the expectations of~\eqref{eq:return_econ} and~\eqref{eq:return_cont}, we get
\begin{equation} 
z_t = -\frac{a_r}{b_r} + (-𝜎^2/2 + 𝜎𝜃)\frac{𝛥t}{b_r} - (1-ⅇ^{-𝜅𝛥t}) \frac{𝜎𝜃}{b_r 𝜅} + (1-ⅇ^{-𝜅𝛥t}) \frac{𝜎}{b_r 𝜅} X_{t_n}
\label{eq:z_demo}
\end{equation} 
Iterating forward~\eqref{eq:z_demo}, $𝛥t$ periods ahead and using~\eqref{eq:dp_econ}, we obtain 
\begin{align}
-\frac{a_r}{b_r} + (-𝜎^2/2 + 𝜎𝜃)\frac{𝛥t}{b_r} - (1-ⅇ^{-𝜅𝛥t}) \frac{𝜎𝜃}{b_r 𝜅} + (1-ⅇ^{-𝜅𝛥t}) \frac{𝜎}{b_r 𝜅} X╻ = {} \nonumber\\
 a_z + b_z \left(-(-𝜎^2/2 + 𝜎𝜃)\frac{𝛥t}{b_r} + (1-ⅇ^{-𝜅𝛥t}) \frac{𝜎𝜃}{b_r 𝜅} - (1-ⅇ^{-𝜅𝛥t}) \frac{𝜎}{b_r 𝜅} X_{t_n}\right) + 𝜀_{t+𝛥t}^z
\end{align}
After some algebra, we find that
\begin{align}
X╻ = & \left[ -\frac{a_z b_r}{𝜎}+(1{-}b_z)\left(-\frac{a_r}{𝜎} + (-𝜎/2+𝜃)𝛥t-𝜃\right)\right] \frac{𝜅}{1-ⅇ^{-𝜅𝛥t}} + b_z X_{t_n} - {} \nonumber \\
&\frac{b_r}{𝜎} \frac{𝜅}{1-ⅇ^{-𝜅𝛥t}} 𝜀_{t+𝛥t}^z
\label{eq:X_demo}
\end{align}
Notice that~$\lim_{𝜅𝛥t \to 0} \left(1-ⅇ^{-𝜅𝛥t}\right) = 𝜅𝛥t$. Finally, comparing equation~\eqref{eq:X_demo} to~\eqref{eq:X_cont}, equations~\eqref{eq:r}-\eqref{eq:kappa} directly follow. To compute the associated second moments, one can compute the variance of~$U$ vector in~\eqref{eq:u_matrix}.
\begin{equation}
\text{Var}
\begin{pmatrix}
U^p╻ \\ U^x╻
\end{pmatrix}
= \int_{𝜏=0}^{𝛥t} FF'
\begin{pmatrix}
ⅆ 𝜏 \\ ⅆ 𝜏
\end{pmatrix}
\label{eq:variance_U}
\end{equation}
where
\begin{equation}
F =
\begin{pmatrix}
1 & (1-ⅇ^{-𝜅𝛥t}) \frac{𝜎}{𝜅} \\
0 & ⅇ^{-𝜅𝛥t}
\end{pmatrix}
\begin{pmatrix}
𝜎 & 0\\
𝜁𝜌 & 𝜁\sqrt{1{-}𝜌^2}
\end{pmatrix}
\end{equation}
Using a matching procedure involving equations~\eqref{eq:return_econ}-\eqref{eq:variance_U} we
 can directly reset equations~\eqref{eq-sigma}-\eqref{eq-rho}. Furthermore, the resulting general formula for every~$𝛥t$ become such that
 \begin{align}
 \text{Var}(X╻) = & \frac{𝜁^2}{2𝜅}(1-ⅇ^{-2𝜅𝛥t}) \\
\text{Cov}(X╻,𝛥\log P_{t+𝛥t}) = & \frac{\rho𝜎𝜁}{𝜅}\left(1-ⅇ^{-𝜅𝛥t}\right)+\frac{𝜎𝜁^2}{𝜅^2}\left(1-ⅇ^{-𝜅𝛥t}\right)-\frac{𝜎𝜁^2}{2𝜅^2}\left(1-ⅇ^{-2𝜅𝛥t}\right) \\
\text{Var} \left(𝛥 \ln ({P}_{t+𝛥t})\right) = & \left(𝜎^2+2\rho\frac{𝜁 𝜎^2}{𝜅}+\frac{𝜁^2 𝜎^2}{𝜅^2}\right)𝛥t-2\rho\frac{𝜁 𝜎^2}{𝜅^2}\left(1-ⅇ^{-𝜅(𝛥t)}\right) \nonumber\\
 & {}-2\frac{𝜁^2 𝜎^2}{𝜅^{3}}\left(1-ⅇ^{-𝜅(𝛥t)}\right)+\frac{𝜁^2𝜎^2}{2𝜅^{3}}\left(1-ⅇ^{-2𝜅(𝛥t)}\right) 
\end{align}
Where the instantaneous standard deviation of~$X$ denoted~$𝜁$ is given by equation~\eqref{eq:Sigma_theta}. Again, notice that, for small~$𝜅$, i.e. when~$𝜅𝛥t \rightarrow 0$, the term~$(1-ⅇ^{-𝜅𝛥t}) \rightarrow 𝜅𝛥t$. So when~$𝛥t=1$, all second moments could be approximated by their instantaneous counterparts. Otherwise, when~$𝛥t \ne 1$, these kinds of approximations become no longer valid. \citet[p.~2208]{Campbell-al.-2004} discuss about pitfalls resulting for this case. Taking this into account, for instance, one can compute the terminal conditional variances by just setting~$𝛥t= T$ and~$t=0$.

The unconditional moments of~$X$ that have been used in this paper are derived from equation~\eqref{eq:z_demo} when~$𝛥t$ is normalized to one.
\begin{align}
X_{t_n} = & \frac{𝜎}{2}+\frac{a_r+b_rz_t}{𝜎}\\
\text{E}(X_{t_n}) = & \frac{𝜎}{2} + \frac{a_r + b_r \text{E}(z_t)}{𝜎} = \frac{𝜎}{2} + \frac{a_r + b_r a_z/(1{-}b_z)}{𝜎}
\end{align}
Then, the unconditional mean of~$X$ is
\begin{equation}
𝜃 = \frac{a_z b_r}{𝜎_r(1{-}b_z)} + \frac{a_r + 𝜎_r^2/2}{𝜎_r}
\end{equation}
In fact, we have used the result~$𝜎=𝜎_r$ in equation~\eqref{eq-sigma} and the fact that~$z$ follows an AR(1) process (\cite{Brandt-al.-2005} followed by \citet{Garlappi-Skoulakis-2009} among others). Thus its unconditional moments are known, $\text{E}\left(z_t\right)=a_z/(1-b_z)$ and~$\text{Var}\left(z_t\right)=𝜎^2_z/(1-b^2_z)$ (\citet[p.~53]{Hamilton-1994}). So, under equation~\eqref{eq:z_demo}, one can match all unconditional percentiles~$z_{(p)}$ with their unconditional counterparts~$X_{(p)}$ ($p$ denotes the~$p$-th percentile) and get the optimal policies for those values.  Since the hedging demand is very sensitive to state variable, we directly draw~$X_{(p)}$ from the unconditional distribution of the point observation~$X_{t_n}$ of our continuous time state variable~$X$ in order to avoid computational errors. As a result, $X_{(50)} = 𝜃$ and the following unconditional distributions hold.
\begin{equation}
z \sim N\left(\frac{a_z}{1-b_z},\,\frac{𝜎_z^2}{1-b_z^2}\right) \quad \Longrightarrow \quad X \sim N \left(𝜃,\,\frac{b_r^2}{\left(1-b_z^2\right)}\frac{𝜎_z^2}{𝜎_r^2}\right)
\label{eq:Unc-dist}
\end{equation}

\subsection*{A.2 Proof of parameters $C_1$ and $C_2$}

Regarding~\eqref{eq:system_0}, the solution for~$C_2$ could be derived from the following equation: 
\begin{equation}
\frac{ⅆ C_2}{ⅆ t} + a\,C_2^2 + b\,C_2 + c = 0
\end{equation}
This equation could straightforward be rewritten as
\begin{equation}
\int_t^T \frac{1}{a\,C_2^2 + b\,C_2 + c} ⅆ C_2 = -(T-t)
\end{equation}

Since parameters~$a$, $b$ and~$c$ are constant, given~$C_2(T)=0$, the integral table provides the solution for~$C_2$. Substitute this into the following equation 
\begin{equation}
\frac{ⅆ C_1}{ⅆ t} + 𝜅 𝜃 C_2 + \left(\frac{b}{2} + a \, C_2 \right) C_1 = 0
\end{equation}
that we derived from equation~\eqref{eq:system_0}. Again using the terminal condition~$C_1(T)=0$ and the constant variation method, one can get the solution for~$C_1$ as follows : 
\begin{align*}
C_1(t) & = \int_t^T ⅇ^{-\int_{t+s}^T (b/2 + a C_2(u)) ⅆu}\left(-𝜅𝜃 C_2(s)\right)ⅆs\\
 & = \int_t^Te^{\left(-b(T-t-s)/2 - a \int_{t+s}^T C_2(u) ⅆu\right)}\left(-𝜅𝜃 C_2(s)\right)ⅆs\\
 & = \int_t^Te^{-𝛿(T-t-s)/2}\frac{(𝛿+b)(ⅇ^{𝛿 s}-1)+2𝛿}{(𝛿+b)(ⅇ^{𝛿(T-t)}-1)+2𝛿}\left(-𝜅𝜃 C_2(s)\right)ⅆs\\
 & = \frac{2c𝜅𝜃 ⅇ^{𝛿(T-t)/2}}{(𝛿+b)(ⅇ^{𝛿(T-t)}-1)+2𝛿} \int_t^T \left(ⅇ^{𝛿s/2}-ⅇ^{-𝛿s/2}\right)ⅆs\\
 & = \frac{4c𝜅𝜃/𝛿}{-(𝛿+b)(1-ⅇ^{𝛿(T-t)})+2𝛿}\left(ⅇ^{𝛿(T-t)/2}-1\right)^2\\
& =  \frac{2𝜅𝜃}{𝛿}\frac{2c}{2𝛿-(𝛿+b)(ⅇ^{𝛿(T-t)}-1)}\left(1-ⅇ^{𝛿(T-t)/2}\right)^2\\
 & = \frac{2𝜅𝜃}{𝛿}\frac{\left(1-ⅇ^{𝛿(T-t)/2}\right)^2}{1-ⅇ^{𝛿(T-t)}}C_2(t)\\
\end{align*}
\end{document}